\documentclass[12pt]{article}
\usepackage{amsmath,amssymb,epsfig}
\usepackage{cancel}
\usepackage{color}
\usepackage{bm}
\usepackage{braket}

\makeatletter

\@addtoreset{equation}{section}
\makeatother

\begin{document}

\title{\vbox{
\baselineskip 14pt
\hfill \hbox{\normalsize KUNS-2596, EPHOU-15-018,WU-HEP-15-22}
} \vskip 1.7cm
\bf Majorana neutrino mass structure induced by rigid instatons on toroidal orbifold \vskip 0.5cm
}
\author{
Tatsuo~Kobayashi$^{1}$, \ \
Yoshiyuki~Tatsuta$^{2}$, \ and \
Shohei~Uemura$^{3}$
\\*[20pt]
{\it \normalsize 
${}^{1}$Department of Physics, Hokkaido University, 
Sapporo, 060-0810 Japan}
\\
{\it \normalsize 
${}^{2}$Department of Physics, Waseda University, 
Tokyo 169-8555, Japan}
\\
{\it \normalsize 
${}^{3}$Department of Physics, Kyoto University, 
Kyoto 606-8502, Japan}
\\*[50pt]}

\date{
\centerline{\small \bf Abstract}
\begin{minipage}{0.9\linewidth}
\medskip 
\medskip 
\small
We study effects of D-brane instantons  wrapping rigid cycles on $\mathbb{Z}_2\times\mathbb{Z}_2'$ toroidal orbifold.
We compute Majorana masses induced by rigid D-brane instantons and realize bimaximal mixing matrices in certain models.
We can also derive more generic mass matrices in other models.
The bimaximal mixing Majorana mass matrix has a possibility to explain observed mixing angles.
We also compute the $\mu$-term matrix among more than one pairs of Higgs fields induced by rigid D-brane instantons. 
\end{minipage}
}

\newpage

\begin{titlepage}
\maketitle
\thispagestyle{empty}
\clearpage
\end{titlepage}

\renewcommand{\thefootnote}{\arabic{footnote}}
\setcounter{footnote}{0}

\section{Introduction}

The Standard Model (SM) is the most successful theory in particle physics.
However,  a lot of mysteries still remain there.
Quantum gravity theory is the biggest one and superstring theory is the most promising (and almost only) candidate 
for quantum gravity theory.
In addition, superstring theory may be able to unify all interactions and matters, too.
Therefore, much effort has been made to understand particle physics within the framework of superstring theory.
Superstring theory has no dimensionless parameters but numerous perturbative vacua.
The D-brane model building is one of interesting methods to construct explicit vacua because one can simply realize 
gauge symmetry and chiral structure  of the SM 
\cite{Berkooz:1996km,Blumenhagen:2000wh,Aldazabal:2000dg,Angelantonj:2000hi} as well as other aspects.
(See for review \cite{Blumenhagen:2006ci,Ibanez:2012zz}.)
We can construct many semi-realistic models, for example models with the SM gauge symmetry, chiral spectrum 
and the Higgs sector and no other exotics \cite{Cvetic:2001nr,Ibanez:2001nd}.
It is the first step to realize the SM gauge symmetry and chiral generation spectrum.
The next issue to realize the SM would be explanation about more detailed and quantitative aspects, e.g. the origin of hierarchy between electroweek scale and string scale
, the flavor structure, etc.

One obstruction for D-brane models to construct realistic models is that extra $U(1)$ symmetries of D-branes forbid some phenomenologically required terms.
For instance, D-brane models often have right-handed neutrinos in order to cancel RR tadpoles.
From phenomenological interest, heavy Majorana mass terms of right-handed neutrinos  
such as $\mathcal{O}(10^{10}-10^{15})$GeV are favored.
However, extra $U(1)$ symmetries on D-branes often forbid Majorana masses of right-handed neutrinos perturbatively.
In D-brane models, such perturbative symmetries can be violated by non-perturbative effects, i.e., 
Dp-brane instantons (or E-branes), which are Dp-branes localized at points in 4 dimensional Minkowski space and wrapping 
(p+1)-cycles on the compact space \cite{Blumenhagen:2006xt,Blumenhagen:2009qh,Ibanez:2006da,Ibanez:2007rs,Cvetic:2007ku}.\footnote{See also \cite{Hamada:2014hpa}.}

Our purpose in this paper is to study explicitly instanton-induced Majorana masses of right-handed neutrinos 
as well as $\mu$-term matrices within the framework of type IIA orientifold models, in particular 
intersecting D6-brane models on toroidal orbifolds. 
Here, we concentrate on D2-brane instantons (or E2-brane) in IIA orientifold models compactified on 
the $\mathbb{Z}_2\times \mathbb{Z}_2'$ toroidal orbifold.
On the $\mathbb{Z}_2\times \mathbb{Z}_2'$ toroidal orbifold, E2-branes can wrap rigid cycles whose position moduli in the compact space are frozen and the positions are fixed \cite{Blumenhagen:2005tn}. 
Such a D-brane instanton can induce superpotential terms non-perturbatively, i.e. Majorana masses of right-handed neutrinos and 
the Higgs $\mu$-term.
We will compute them explicitly, and show that 
for Majorana neutrino mass matrix, there is a typical $\mathbb{Z}_2$ symmetry inherited from the geometrical symmetry 
of the $\mathbb{Z}_2\times \mathbb{Z}_2'$ orbifold  in certain models.
For the Higgs $\mu$-term matrix among more than one pairs of Higgs fields, 
one D-brane instanton can make only the rank-one $\mu$-term matrix.
However, there are more than 10 rigid cycles on the $\mathbb{Z}_2\times \mathbb{Z}_2'$ orbifold and it may be  possible to make the full rank $\mu$-term matrix.

This paper is organized as follows.
In section 2, we review D2-brane instantons and rigid cycles on toroidal orbifolds.
In section 3, we compute  generic form of Majorana neutrino mass matrix induced by D-brane instantons 
and compute explicitly some illustrating examples.
As a result, we obtain a bimaximal mixing matrix for Majorana neutrino mass matrix in certain models.
This symmetry is inherited from the geometric symmetry of the $\mathbb{Z}_2\times \mathbb{Z}_2'$ orbifold.
In section 4, we also compute Higgs $\mu$-term matrix for more than one pairs of Higgs supermultiplets.
Section 5 is the conclusion.

\section{D-brane instanton and toroidal orbifolds}

In this section, we briefly review D-brane instanton (or E-brane) effects and rigid cycles on toroidal orbifolds.
For more detailed review of E-branes, see \cite{Ibanez:2012zz,Ibanez:2007rs,Cvetic:2007ku}.

\subsection{D-brane instanton}

A Dp-brane instanton is a D-brane localized at a point in 4 dimensional spacetime and wrapping a (p+1)-cycle on the 6 dimensional compact space.
It is similar to an instanton in gauge theory and it is named as D-brane instanton.
In IIA superstring theory, there are E2-branes and we concentrate on them.

Similar to the gauge instanton, a D2-brane instanton has zero-modes $\alpha_i$ and non-perturbative effects of the D-brane instanton are
 computed by integrating these zero-modes,
\begin{equation}
\int \prod_{i} \mathcal{D}\alpha_i  e^{-S_{{\rm DBI}}-S_{{\rm CS}}}e^{-S_{{\rm interact}}},
\end{equation}
where 
 $S_{\rm DBI}$, $S_{{\rm CS}}$ and $S_{{\rm interact}}$ 
denote the Dirac-Born-Infeld action, the Charn-Simons term and the interaction among zero-modes of D2-brane instanton and visible matters on D6-branes, 
respectively.
The Charn-Simons term can be written as
\begin{equation}
S_{{\rm CS}}=   i N_{{\rm E}}\int_{\Pi_{{\rm E}}}C_{3} , 
\end{equation}
where $N_{{\rm E}}$ is the multiplicity of E-branes. Here, $\Pi_{{\rm E}}$ denotes 
the homology class of the cycle which the E-brane wraps.
We introduce $[\alpha_{k}]$ as the basis of 3-cycles and its dual basis $[\beta_{l}]$, where [$\alpha_k]\circ [\beta_l]=\delta_{kl}$.
Also, we define the axion,
\begin{equation}
a_k=\int_{\beta_k} C_3.
\end{equation}
Then, we can write 
\begin{equation}
S_{{\rm CS}}= i\sum_{k}N_{{\rm E}}q_{{\rm E},k} a_{k},
\end{equation}
where $q_{{\rm E},k}=[\Pi_{{\rm E}}]\circ[\beta_k]$.

As mentioned in the previous section, the $U(1)_a$ symmetry of $U(N_a)=U(1)_a\times SU(N_a)$ on the D6$_a$-brane 
forbids perturbatively gauge variant operators ${\cal O}_{c}$, which transform 
\begin{equation}
{\cal O}_{c} \rightarrow e^{iq_c\Lambda_a}{\cal O}_{c},
\label{eq:O-variant}
\end{equation}
 under the $U(1)_a$ transformation $A_\mu^a \rightarrow A_\mu^a +\partial_\mu \Lambda_a$.
However, the axion shifts 
\begin{equation}
a_k \rightarrow a_k + N_a Q_{a,k}\Lambda_a,
\end{equation}
under the  $U(1)_a$ transformation, where 
$ Q_{a,k} = [\alpha_k]\circ [\Pi_a]$.
That leads to the following extra $U(1)_a$ transformation of instanton effects, 
\begin{equation}
e^{-S_{{\rm CS}}}~\rightarrow~e^{-S_{{\rm CS}}}{\rm exp}\left( -i\sum_{k}N_{a}N_{{\rm E}}Q_{a,k}q_{{\rm E},{k}}\Lambda_{a} \right).
\end{equation}
Thus, the E-brane effect can cancel  $U(1)$ phase  in Eq.~(\ref{eq:O-variant}) for certain gauge variant operators, 
and generate interactions, which are  perturbatively forbidden by extra $U(1)$ symmetries.

The zero-modes $\alpha_i$s are classified into two classes; neutral zero-modes and charged zero-modes.
Neutral zero-modes are not charged under gauge groups of other D-branes.
Neutral zero-modes are zero-modes of open string stretching between a D-brane instanton.
These zero-modes correspond to  position moduli of the D-brane instanton and their superpartners.
Charged zero-modes are charged under gauge groups of other D-branes.
These zero-modes are zero-modes of open string stretching between a D-brane instanton and another D-brane.
These zero-modes are similar to chiral superfields.

In order to generate superpotential terms by the D-brane instanton, we need two fermionic neutral zero-modes $\theta$, 
which correspond to the D-brane instanton position on the superspace, and also 
the goldstinos of supersymmety broken by the D-brane instanton.
Then, non-perturbative effects are expressed as
\begin{equation}
\int d^2\theta d^4 x \prod_i \mathcal{D}\alpha_{i}  e^{-S_{{\rm DBI}}-S_{{\rm CS}}}e^{-S_{{\rm interact}}},
\end{equation}
where $\alpha_i$s are charged zero-modes.
Finally we can generate the non-perturbative superpotential term,
\begin{equation}
\mathcal{W}=\prod_i (\psi_{i})^{n_i}  e^{-S_{{\rm DBI}}}.
\end{equation}

If extra neutral zero-modes appear and those are not lifted up, 
the non-perturbatively induced superpotential term vanishes.
If the E-brane can move freely in compact space, it has extra neutral zero-modes.
Then, we should consider E-brane wrapping rigid cycles.

\subsection{Rigid cycles on toroidal orbifolds}

The toroidal orbifolds are the simple examples having rigid cycles.
In Table \ref{tab:Hodge_number}, we list toroidal supersymmetric orbifolds and the Hodge number of them.
In this paper, we follow the notations in \cite{Blumenhagen:2005tn}.
The orbifold twist is denoted by $\Gamma$.
\begin{table}[th]
\begin{center}
\begin{tabular}{c|ccccccccc}
\hline
\hline
   $\Gamma$      & $\mathbb{Z}_3$ & $\mathbb{Z}_4$& $\mathbb{Z}_6$& $\mathbb{Z}_6'$& $\mathbb{Z}_2\times\mathbb{Z}_2$& $\mathbb{Z}_2\times\mathbb{Z}_2'$& $\mathbb{Z}_2\times\mathbb{Z}_4$& $\mathbb{Z}_3\times\mathbb{Z}_3$& $\mathbb{Z}_3\times \mathbb{Z}_6$\\
\hline
$h_{11}^{{\rm unt}}$ & 9 &5 & 5 & 3 & 3&3 & 3 & 3 &3 \\
$h_{11}^{{\rm twt}}$ &27 & 26  & 24 & 32 & 48 & 0 & 58 & 81 & 70\\
\hline
$h_{21}^{{\rm unt}}$ & 0  & 1   & 0 & 1 & 3 & 3 & 1 & 0 & 0\\
$h_{21}^{{\rm twt}}$ & 0 & 6  & 5 & 10 &0 & 48 &0 & 0& 1\\
\hline
\hline
\end{tabular}
\end{center}
\caption{The Hodge numbers of the fractional toroidal orbifold ($T^2\times T^2\times T^2)/\Gamma$ having $\mathcal{N}=1$ SUSY.}
\label{tab:Hodge_number}
\end{table}

For type IIA intersecting D6-brane models, non-pertubative effects originate from D2-brane instanton wrapping 3-cycles on compact space.
The number of independent 3-cycles is given by the Betti number, $b_3 = 2+2h_{21}^{{\rm unt}}+2h_{21}^{{\rm twt}} $.
The bulk 3-cycles are inherited from 3-cycles on the covering torus $T^6$,
\begin{equation}
\Pi^B_a=\sum_{g\in\Gamma} \mathcal{R}_{g}\cdot \Pi_a^{T^6} ,
\end{equation}
where $\Pi_a^{T^6}$ is the $a$-th 3-cycle on $T^6$ and $\mathcal{R}_g$ is the geometrical action of $g\in\Gamma$. 
The number of these cycles is 2+2$h_{21}^{{\rm unt}}$ and these cycles have position moduli.

The number of fractional (or twisted) 3-cycles is given by $2h_{21}^{{\rm twt}}$.
These cycles originate from orbifold fixed points and can not move away from the fixed points.
Then, they have no position moduli and zero-modes.

There are five toroidal orbifolds having rigid cycles as shown in Table \ref{tab:Hodge_number}.
Here, we concentrate on the $\mathbb{Z}_2\times\mathbb{Z}_2'$ orbifold model
among these five torodal orbifolds, because it has the most rigid cycles and rigid cycles on the other orbifolds have self intersection numbers.
Hence, the other rigid cycles have extra neutral zero-modes.
At any rate, the $\mathbb{Z}_2\times\mathbb{Z}_2'$ orbifold is the most simple example.

\subsection{$\mathbb{Z}_2\times\mathbb{Z}_2'$ toroidal orbifold}

Homology classes of rigid 3-cycles  on the $\mathbb{Z}_2\times\mathbb{Z}_2'$ orbifold are given by
\begin{equation}
\Pi_a^F=\frac{1}{4}\Pi_a^B+\frac{1}{4}\left(\sum_{i,j\in S^a_{\Theta}} \epsilon^{\Theta}_{a,ij} \Pi_{ij,a}^{\Theta} \right)
+\frac{1}{4}\left(\sum_{j,k\in S^a_{\Theta'}} \epsilon^{\Theta'}_{a,jk} \Pi_{jk,a}^{\Theta'} \right)
+\frac{1}{4}\left(\sum_{i,k\in S^a_{\Theta \Theta'}} \epsilon^{\Theta \Theta'}_{a,ik} \Pi_{ik,a}^{\Theta \Theta'} \right).
\end{equation}
Here, $\Theta$ and $\Theta'$ are the generators of $\mathbb{Z}_2$ and $\mathbb{Z}_2'$, respectively.
That is, $\Theta$ acts the complex coordinates $z_i$ on the $i$-th $T^2$ as 
$(z_1,z_2,z_3) \rightarrow (-z_1,-z_2,z_3)$, while $\Theta'$ acts  as 
$(z_1,z_2,z_3) \rightarrow (z_1,-z_2,-z_3)$.
Also, $S_g^a$ denotes the set 
of the fixed points of  $\mathcal{R}_g$ which D6$_a$-brane wraps.
In addition, $i,j,k$ are the index numbers corresponding to the fixed points on the first, second and third $T^2$ respectively.
All of $i,j,k$ vary from 1 to 4 and one of them represents one of four fixed points on each torus.
$\epsilon^{g}_{a,ij}$ corresponds to the two possible orientations of the collapsed 2-cycles on the fixed point of 
 $\mathcal{R}_g$, i.e. $\epsilon^{g}_{a,ij} = \pm 1$.
$\Pi_{ij}^g$ is the collapsed 3-cycle which can be given by
\begin{equation}
[\Pi_{ij}^{g}]=n_a^{I_g}[\alpha_{ij,n}^g]+\tilde{m}_a^{I_g}[\alpha_{ij,m}^g],
\end{equation}
where $[\alpha_{ij,n}^g]$ is the product of the collapsed 2-cycle on the $i$-th and $j$-th fixed points of $\mathcal{R}_g$ and the 1-cycle on the other torus.
The number of chiral zero-modes between Dp-branes wrapping $\Pi_{a}^F$ and $\Pi_{b}^F$ is given by the intersection number of rigid 3-cycles.
The intersection number of rigid 3-cycles is obtained as \cite{Blumenhagen:2005tn}
\begin{equation}
\begin{split}
I_{ab}^F&=\Pi_a^F \cdot \Pi_b^F,\\
          &= \frac{1}{4} I_{ab}^{{\bf T}^6}+\frac{1}{4}\sum_{g\in G} \sum_{i,j\in S_g^a} \sum_{k,l \in S_g^b} \epsilon_{a,ij}^g \epsilon_{b,kl}^g \delta_{ik}\delta_{jl} (n_a^{I_{g}} \tilde{m}_b^{I_g} - \tilde{m}_a^{I_g} n_b^{I_g}),\\
&= \frac{1}{4} I_{ab}^{{\bf T}^6}+\frac{1}{4}\sum_{g\in G} \sum_{i,j\in S_g^a} \sum_{k,l \in S_g^b} \epsilon_{a,ij}^g \epsilon_{b,kl}^g \delta_{ik}\delta_{jl} I_{ab}^{I_g}.
\end{split}
\label{eq:intersection_number}
\end{equation}

We need to know physical states on the $\mathbb{Z}_2 \times \mathbb{Z}_2'$ orbifold, but not the total intersection number, in order to compute 
explicitly couplings of states and mass terms.
The above number of these zero-modes is interpreted as the result of projection of zero-modes on the covering space $T^6$.
The number is equal to the number of zero-modes invariant under the action of $\mathbb{Z}_2\times \mathbb{Z}_2'$. 
We can write $\mathbb{Z}_2\times \mathbb{Z}_2'$ invariant states as follows
\begin{equation}
\psi_{{\rm orbifold},i} = \frac{1}{C_{\psi_i}}\left( \psi_i + \Delta_{\Theta}^{\psi} \psi_{\Theta i} +\Delta_{\Theta'}^{\psi} \psi_{\Theta' i} + \Delta_{\Theta \Theta'}^{\psi}\psi_{\Theta \Theta' i}\right),
\label{eq:invariant_state}
\end{equation}
where $\psi_i$ is the state localized at the $i$-th intersection point and $C_{\psi_i}$ is the normalization factor of the state.
For instance, if $\psi_i$ is invariant under $\Theta$ and $\Theta'$, $C_{\psi_i}=4$.
$g\in \mathbb{Z}_2\times\mathbb{Z}_2'$ transforms an $i$-th intersection point to another point $g(i)$.
$\Delta_{g}^{\psi}$ is the phase of the action of $g\in \mathbb{Z}_2\times \mathbb{Z}_2'$ determined by the sign of intersecting numbers.\footnote{$\Delta_{g}^{\psi}$ corresponds to the generalized GSO phase in 
heterotic orbifold models. By the GSO projection, we can obtain the twist invariant states \cite{Kobayashi:1990mc,Kobayashi:2004ya} , 
and also the total number of massless modes by inserting it in the partition function \cite{Ibanez:1987pj,Senda:1987pf}.}
Similar $\mathbb{Z}_N$ eigenstates are discussed in magnetized brane models \cite{Abe:2008fi,Abe:2013bca}.
For instance, if all of $\epsilon_{a,ij}^g$ and $I_{ab}^{I_g}$ are positive, these phases $\Delta_{g}^\psi$ are 
equal to $\Delta_{g}^\psi =1$.
Then, we obtain $ I_{ab}^F$ zero-modes as 
\begin{equation} 
 I_{ab}^F = \frac{1}{4} I_{ab}^{{\bf T}^6}+\frac{1}{4}\sum_{g\in G} \sum_{i,j\in S_g^a} \sum_{k,l \in S_g^b} \delta_{ik} \delta_{jl} I_{ab}^{I_g}.
 \end{equation}

\section{Right-handed neutrino Majorana masses}

Here we study right-handed neutrino Majorana masses on the $\mathbb{Z}_2\times\mathbb{Z}_2'$ toroidal orbifold.
We concentrate on three generations of right-handed neutrinos $N_R^i $.
They are localized at the intersection points between D6$_a$-brane and D6$_b$-brane wrapping rigid cycles.
The Majorana masses are perturbatively forbidden by the $U(1)$ symmetries of D6-branes.
However, they can be generated by D2-brane instanton effects.
A D2-brane instanton intersects with D6$_{a}$-brane and D6$_{b}$-brane. Open strings at these intersection points have zero-modes $\alpha_i$ and $\beta_j$ respectively.
Only if there are two zero-modes and they have couplings such as $d^{ij}_a \alpha_i N_R^a \beta_j$ with 
the 3-point coupling $d^{ij}_a$, Majorana neutrino masses are generated as
\begin{eqnarray}
& & M\int d^2\alpha d^2 \beta ~e^{-d^{ij}_a \alpha_i N_R^a \beta_j} 
= \sum_{a,b} M N^a_R N^b_R  c_{ab}, \nonumber \\
& & c_{ab} = 
\epsilon_{ij} \epsilon_{k \ell}d^{ik}_a d^{j \ell}_b,
\end{eqnarray}
where $M$ is determined by the string scale $M_s$ and the volume of E-branes $V$ like $M=M_s e^{-V}$.

The 3-point coupling $d_{ij}^a$ is given by the linear combination of 3-point couplings on the covering torus.
Because $\mathbb{Z}_2 \times \mathbb{Z}_2'$ invariant states are given by eq (\ref{eq:invariant_state}), the 3-point coupling of these states is computed as
\begin{equation}
d_a^{ij}=\frac{1}{C_{N_R^a}C_{\alpha_i}C_{\beta_j}}\sum_{f,g,h\in\mathbb{Z}_2\times\mathbb{Z}_2'} \Delta_f^{N_R} \Delta_g^\alpha \Delta_h^\beta y_{f(a)}^{g(i) h(j)},\\
\end{equation}
where $y_{a}^{ij}$ is the 3-point coupling on covering $T^6$.
The 3-point coupling on $T^6$ is obtained by worldsheet instanton.
The intersection number on covering $T^6$ can be decomposed as $I_{ab}=I_{ab}^1I_{ab}^2I_{ab}^3$, 
where $I_{ab}^n$ denotes the intersection number on the $n$-th $T^2$.
The 3-point coupling is also decomposed as
\begin{equation}
y_a^{ij}=y_{a,1}^{ij}y_{a,2}^{ij}y_{a,3}^{ij},
\end{equation}
where $y_{a,n}^{ij}$ is the 3-point couplings on the $n$-th  $T^2$. 
By using the $\vartheta$-function, the 3-point coupling $y_{a,n}^{ij}$ is given as follows
\cite{Cvetic:2003ch}\footnote{Similar results for 3-point couplings and higher order couplings are 
obtained  
in magnetized brane models and heterotic orbifold models \cite{Cremades:2004wa,Abe:2009dr,Hamidi:1986vh,Abe:2015yva}.}
\begin{equation}
y_{a,n}^{ij} = C \vartheta \left[
\begin{array}{c}
\frac{a}{I_{ab}^n} + \frac{i}{I_{bc}^n} + \frac{j}{I_{ca}^n}+\frac{I_{bc}^n \varepsilon_a^n + I_{ca}^n \varepsilon_b^n+I_{ab}^n\varepsilon_c^n}{I_{ab}^n I_{bc}^n I_{ca}^n } \\
0  \\
\end{array}
\right] \left( 0,\frac{iA^n | I_{ab}^n I_{bc}^n I_{ca}^n |}{4\pi^{2}\alpha'} \right) ,
\label{eq:yukawa_torus}
\end{equation}
where $A^n$ is the area of the $n$-th $T^2$, and 
$\varepsilon_x^n$ with $x=a,b,c$ is the position moduli of D6$_x$-brane on the $n$-th torus.
In our computations, these moduli are discretized because D-branes wrap rigid cycles.
Since we consider D-branes wrapping rigid cycles, the configuration of branes on covering $T^6$ is invariant under the action of $\mathbb{Z}_2\times\mathbb{Z}_2'$ and 3-point couplings are also invariant, which means $y_{a}^{ij}=y_{g(a)}^{g(i)g(j)}$.
Then, we obtain
\begin{equation}
\begin{split}
d_a^{ij}&=\frac{1}{C_{N_R^a}C_{\alpha_i}C_{\beta_j}}\sum_{f,g,h\in\mathbb{Z}_2\times\mathbb{Z}_2'} \Delta_f^{N_R} \Delta_g^\alpha \Delta_h^\beta y_{a}^{f^{-1}\cdot g(i) f^{-1}\cdot h(j)},\\
&=\begin{cases}
\frac{4}{C_{N_R^a}C_{\alpha_i}C_{\beta_j}}\sum_{g,h\in\mathbb{Z}_2\times\mathbb{Z}_2'} \Delta_g^\alpha \Delta_h^\beta y_{a}^{g(i) h(j)} & (\Delta_f^{N_R}\Delta_f^{\alpha}\Delta_f^{\beta}=1~{\rm for} ~\forall f ),\\
0 &({\rm otherwise}).
\end{cases}
\end{split}
\end{equation}
As a result we can derive the Majorana masses by one instanton configuration
\begin{equation}
c_{ab}=
\begin{cases}
\epsilon_{ij} \epsilon_{k l} \frac{16}{C_{N_R^a}C_{N_R^b}C_{\alpha_i}C_{\beta_k}C_{\alpha_j}C_{\beta_l}}\sum_{g,h,g',h'\in\mathbb{Z}_2\times\mathbb{Z}_2'} \Delta_{g\cdot g'}^\alpha \Delta_{h\cdot h}^\beta y_{a}^{g(i) h(k)} y_{b}^{g'(j) h'(l)} & (\Delta_f^{N_R}\Delta_f^{\alpha}\Delta_f^{\beta}=1~{\rm for}~\forall f ),\\
0 &({\rm otherwise}).
\end{cases}
\end{equation}

If there is another D2-brane instanton, Majorana masses are the sum of these instantons effects, i.e.
\begin{equation}
M_{ab}=\sum_{m} M_m \epsilon_{ij} \epsilon_{k \ell}d^{ik,m}_a d^{j \ell,m}_b.
\end{equation} 
We must take into account all the possible configurations of E-branes.

The 3-point couplings on covering torus (\ref{eq:yukawa_torus}) are determined by the intersection numbers on the torus.
Although we assume the generation number of neutrinos to be equal to three and determine the number of zero-modes, the intersection number on covering torus is not uniquely determined because the intersection number on orbifold (\ref{eq:intersection_number}) depends on the choice of fixed points.
It is difficult to consider all the possibilities, although we would study them systematically elsewhere.
In this paper, we compute some explicit examples to conjecture general form of Majorana masses. 

In toroidal orientifolds, there are two types of $T^2$: the rectangular torus whose complex structure Re$\tau=0$ and 
the tilted torus with Re$\tau$=1/2.
In our analysis, we concentrate on the rectangular torus for simplicity, but  this limitation dose not affect our results.

We use the notation $(l,m,n)_{xy}$ as the abbreviation for the number of fixed points shared by D6$_x$-brane and D6$_y$-brane, 
where $l$ represents the number of fixed points shared by D6$_x$-brane and D6$_y$-brane on the first $T^2$, and $m$ and $n$ are those on the second 
and third $T^2$s.
For example, $(l,m,n)_{xx}$ is always $(2,2,2)_{xx}$.

\subsection{Explicit Model 1}
\label{subsec:ex}

In this section, we compute Majorana neutrino masses in an explicit example.
We consider only D6-branes relating with right-handed neutrinos and E-brane, but do not study complete model building.
We compute the model with $(1,0,0)_{ab}, (1,1,1)_{Ea}$ and $(1,1,1)_{Eb}$, because this is the simplest model with three generations of neutrinos.
In this model, the intersection number is
\begin{equation}
I_{ab}^F=\frac{1}{4}I_{ab}^{1} I_{ab}^{2} I_{ab}^{3}.
\end{equation}
There are only two independent solutions in this model,
\begin{equation}
\begin{cases}
I_{ab}^{1}=1,~~I_{ab}^{2}=6,~~I_{ab}^{3}=2.~~&{\rm (I)}\\
I_{ab}^{1}=3,~~I_{ab}^{2}=2,~~I_{ab}^{3}=2,~~&{\rm (II)}
\end{cases}
\end{equation}
In these models,  there are 12 (would-be neutirino) states on covering torus.
We name these (would-be neutirno) states $\ket{ijk}_{\nu}$, where 
$i,j$ and $k$ represent the index number of fixed points on the first, second and third torus respectively.

For Model 1-(I), we have $i= 0.~j=0,\cdots 5$ and $k=0,1$. 
These states transform under $\mathbb{Z}_2$ and $\mathbb{Z}_2'$ as
\begin{equation}
\begin{split}
\mathbb{Z}_2&: \ket{0jk}_\nu\rightarrow \ket{0(5-j)k}_\nu,\\
\mathbb{Z}_2'&: \ket{0jk}_\nu\rightarrow \ket{0(5-j)(1-k)}_\nu.
\end{split}
\end{equation}
Then, we can write three invariant neutrino states,
\begin{equation}
\begin{split}
N_{0}=\frac{1}{\sqrt{4}}\left(\ket{000}_\nu+\ket{050}_\nu+\ket{051}_\nu+\ket{001}_\nu \right),\\
N_{1}=\frac{1}{\sqrt{4}}\left(\ket{010}_\nu+\ket{040}_\nu+\ket{041}_\nu+\ket{011}_\nu \right),\\
N_{2}=\frac{1}{\sqrt{4}}\left(\ket{020}_\nu+\ket{030}_\nu+\ket{031}_\nu+\ket{021}_\nu \right).
\end{split}
\label{eq:neutrino2}
\end{equation}
For Model 1-(II), we can write three  invariant neutrino states similarly
\begin{equation}
\begin{split}
N_{0}=\frac{1}{\sqrt{4}}\left(\ket{000}_\nu+\ket{010}_\nu+\ket{011}_\nu+\ket{001}_\nu \right),\\
N_{1}=\frac{1}{\sqrt{4}}\left(\ket{100}_\nu+\ket{210}_\nu+\ket{111}_\nu+\ket{201}_\nu \right),\\
N_{2}=\frac{1}{\sqrt{4}}\left(\ket{200}_\nu+\ket{110}_\nu+\ket{211}_\nu+\ket{101}_\nu \right).
\end{split}
\label{eq:neutrino1}
\end{equation}

The intersection number between D2-brane instanton and D6$_a$-brane must satisfy the following condition,
\begin{equation}
I_{Ea}^F=\frac{1}{4}I_{Ea}^{1} I_{Ea}^{2} I_{Ea}^{3}+\frac{1}{4} I_{Ea}^{3}+ \frac{1}{4} I_{Ea}^{1} +\frac{1}{4} I_{Ea}^{2}=2.
\label{eq:alpha}
\end{equation}
There is only one independent solution $(I_{Ea}^{1},I_{Ea}^{2},I_{Ea}^{3})=(\underline{3,1,1})$, where the underline denotes all the possible permutations.
The physical invariant zero-mode states $\alpha_i$ can be written as
\begin{equation}
\begin{split}
\alpha_0&=\ket{000}_{\alpha},\\
\alpha_1&=\frac{1}{\sqrt{2}}\left( \ket{100}_{\alpha} +\ket{200}_{\alpha} \right),
\end{split}
\label{eq:zeros1}
\end{equation}
for the model with $(I_{Ea}^1,I_{Ea}^2,I_{Ea}^3)= (3,1,1)$ by using (would-be zero-mode) states $\ket{ijk}_{\alpha}$.
Similarly, we can write the invariant states $\beta_i$, i.e.
\begin{equation}
\begin{split}
\beta_0&=\ket{000}_{\beta},\\
\beta_1&=\frac{1}{\sqrt{2}}\left( \ket{100}_{\beta} +\ket{200}_{\beta} \right).
\end{split}
\label{eq:zeros1-beta}
\end{equation}

When the three generations of neutrinos (\ref{eq:neutrino2}) or (\ref{eq:neutrino1}) and instanton zero-modes (\ref{eq:zeros1}) appear from different tori, induced Majorana masses vanish \cite{Hamada:2014hpa}.
Therefore, there are only two solutions leading to non-vanishing Majorana masses.
One is the model with $(I_{ab}^{1},I_{ab}^{2},I_{ab}^{3})=(1,6,2)$ and 
$(I_{Ea}^{1},I_{Ea}^{2},I_{Ea}^{3})=(I_{Eb}^{1},I_{Eb}^{2},I_{Eb}^{3})=(1,3,1)$. The other is  $(I_{ab}^{1},I_{ab}^{2},I_{ab}^{3})=(3,2,2)$ and $(I_{Ea}^{1},I_{Ea}^{2},I_{Ea}^{3})=(I_{Eb}^{1},I_{Eb}^{2},I_{Eb}^{3})=(3,1,1)$.
Other solutions have vanishing Majorana masses because it is canceled by the completely antisymmetric tensor $\epsilon_{ab}$.

The interaction term of Model 1-(I) is written as follows\footnote{For more detail, see Appendix \ref{app:yukawa}.}
\begin{equation}
\begin{split}
S_{{\rm interact}}=\sqrt{2} y^1 y_2^2 y^3 \alpha_0 \beta_1 N_0 +\sqrt{2} y^1 y_0^2 y^3 \alpha_1 \beta_0 N_0 +  y^1 y_1^2 y^3 \alpha_1 \beta_1 N_0\\
+\sqrt{2} y^1 y_0^2 y^3 \alpha_0 \beta_1 N_2 + \sqrt{2} y^1 y_2^2 y^3 \alpha_1\beta_0 N_2 +y^1 y_1^2 y^3 \alpha_1 \beta_1 N_2\\
+2 y^1 y_1^2 y^3 \alpha_0 \beta_0 N_1 +y^1 (y_0^2+y_2^2) y^3 \alpha_1 \beta_1 N_1,
\end{split}
\label{eq:interact}
\end{equation}
where $y_i^j$ is the 3-point coupling on the $j$-th torus written as (\ref{eq:yukawa_torus}).
These interactions are determined by $\mathbb{Z}_3$ symmetries of the second torus.
We can derive the following Majorana masses 
\begin{eqnarray}
M_{R,ij} &\propto& (y^1 y^3)^2\left(
\begin{array}{ccc}
2 y^2_0 y^2_2  &  -(y_1^2)^2 &  (y_0^2)^2 + (y_2^2)^2\\
- (y_1^2)^2 & -2 y_1^2(y_0^2+y_2^2) & - (y_1^2)^2\\
 (y_0^2)^2 + (y_2^2)^2 & - (y_1^2)^2 & 2 y^2_0 y^2_2\\
\end{array}
\right)\\
&=&\left(
\begin{array}{ccc}
A & B & C\\
B & D & B\\
C & B & A\\
\end{array}
\right).
\label{eq:bimaximal2}
\end{eqnarray}
Similarly, the Majorana masses of Model 1-(II) are computed as
\begin{eqnarray}
M_{R,ij} \propto (y^2 y^3)^2\left(
\begin{array}{ccc}
y^1_0 y^1_1  & -\frac{1}{4} (y_0^1)^2 & -\frac{1}{4} (y_0^1)^2\\
-\frac{1}{4} (y_0^1)^2 & \frac{1}{2}(y_1^1)^2 & \frac{1}{2}(y_1^1 )^2\\
-\frac{1}{4} (y_0^1)^2 & \frac{1}{2}(y_1^1)^2 & \frac{1}{2}(y_1^1)^2\\
\end{array}
\right)
=\left(
\begin{array}{ccc}
A & B & B\\
B & C & C\\
B & C & C\\
\end{array}
\right).
\label{eq:bimaximal1}
\end{eqnarray}

In these models, apparent $\mathbb{Z}_2$ symmetric (or bimaximal mixing) Majorana mass matrices appear.

\subsection{Explicit model 2}

We study another model.
We consider the model with $(1,1,0)_{ab},(1,1,1)_{Ea}$ and $(0,1,1)_{Eb}$.
In this model, since the E-a intersection number and shared fixed points are the same as the last model, the physical states $\alpha_i$s are the same as eq.(\ref{eq:zeros1}).
The a-b intersection and E-b intersection conditions have a solution
\begin{equation}
\begin{split}
I_{ab}^1=5,~I_{ab}^2=1,~I_{ab}^3=2,\\
I_{Eb}^1=4,~I_{Eb}^2=I_{Eb}^3=1,
\end{split}
\end{equation}
and the physical states of neutrinos are written as
\begin{equation}
\begin{split}
&N_{0}=\frac{1}{\sqrt{2}}\left(\ket{000}_\nu+\ket{001}_\nu\right),\\
&N_{1}=\frac{1}{\sqrt{4}}\left(\ket{100}_\nu+\ket{400}_\nu+\ket{101}_\nu+\ket{401}_\nu \right),\\
&N_{2}=\frac{1}{\sqrt{4}}\left(\ket{200}_\nu+\ket{300}_\nu+\ket{201}_\nu+\ket{301}_\nu \right).
\end{split}
\label{eq:neutrino3}
\end{equation}
The zero-modes $\beta_i$s are 
\begin{equation}
\begin{split}
\beta_{0}=\frac{1}{\sqrt{2}}\left(\ket{000}_\beta+\ket{300}_\beta \right),\\
\beta_{1}=\frac{1}{\sqrt{2}}\left(\ket{100}_{\beta}+\ket{200}_\beta \right).\\
\end{split}
\label{eq:zeros2}
\end{equation}
Then, we can obtain non-vanishing Majorana masses.
In this model, the Majorana mass matrix has no apparent symmetry without fine-tunings, 
but corresponds to  general symmetric Majorana masses with the full rank.

\subsection{Flavor structure of Majorana mass matrix}

We computed Majorana masses in explicit models.
In Model 1, we derive bimaximal mixing mass matrix, and  the other model does not have such a symmetry.
This difference originates from the geometric symmetry of D-brane configurations on the torus.

\begin{figure}[thbp]
\begin{tabular}{cc}
\begin{minipage}{0.5\hsize}
\centering
  \epsfig{file=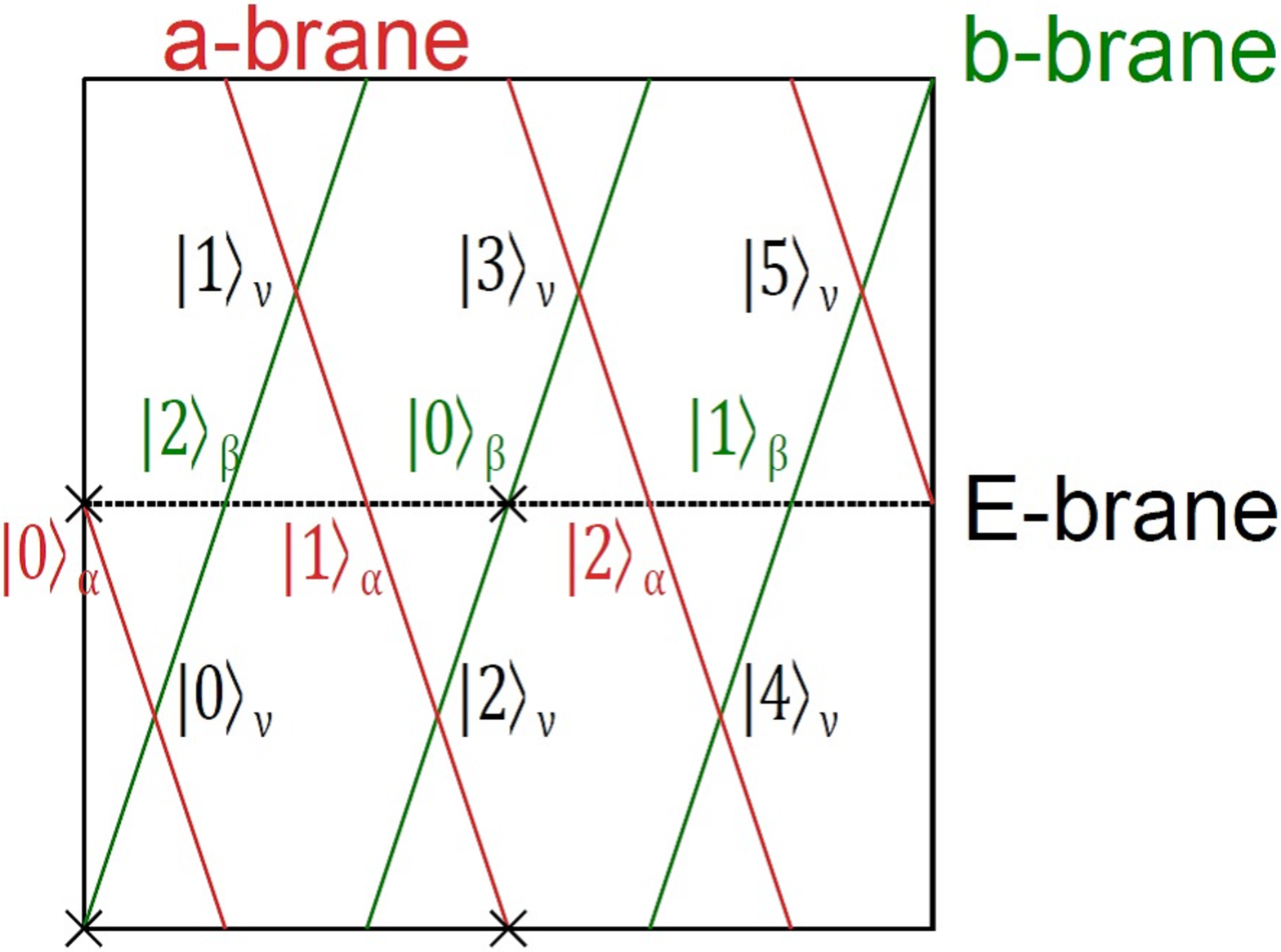,scale=0.25}
 \caption{The brane configuration on the second torus of Model 1-(I).
We omit the index number of other tori, e.g. $\ket{0}_\alpha$ denotes $\ket{000}_\alpha$, the others are similar.}
\label{fig:ex1}
\end{minipage}
\hspace{0.2cm}
\begin{minipage}{0.5\hsize}
\centering
  \epsfig{file=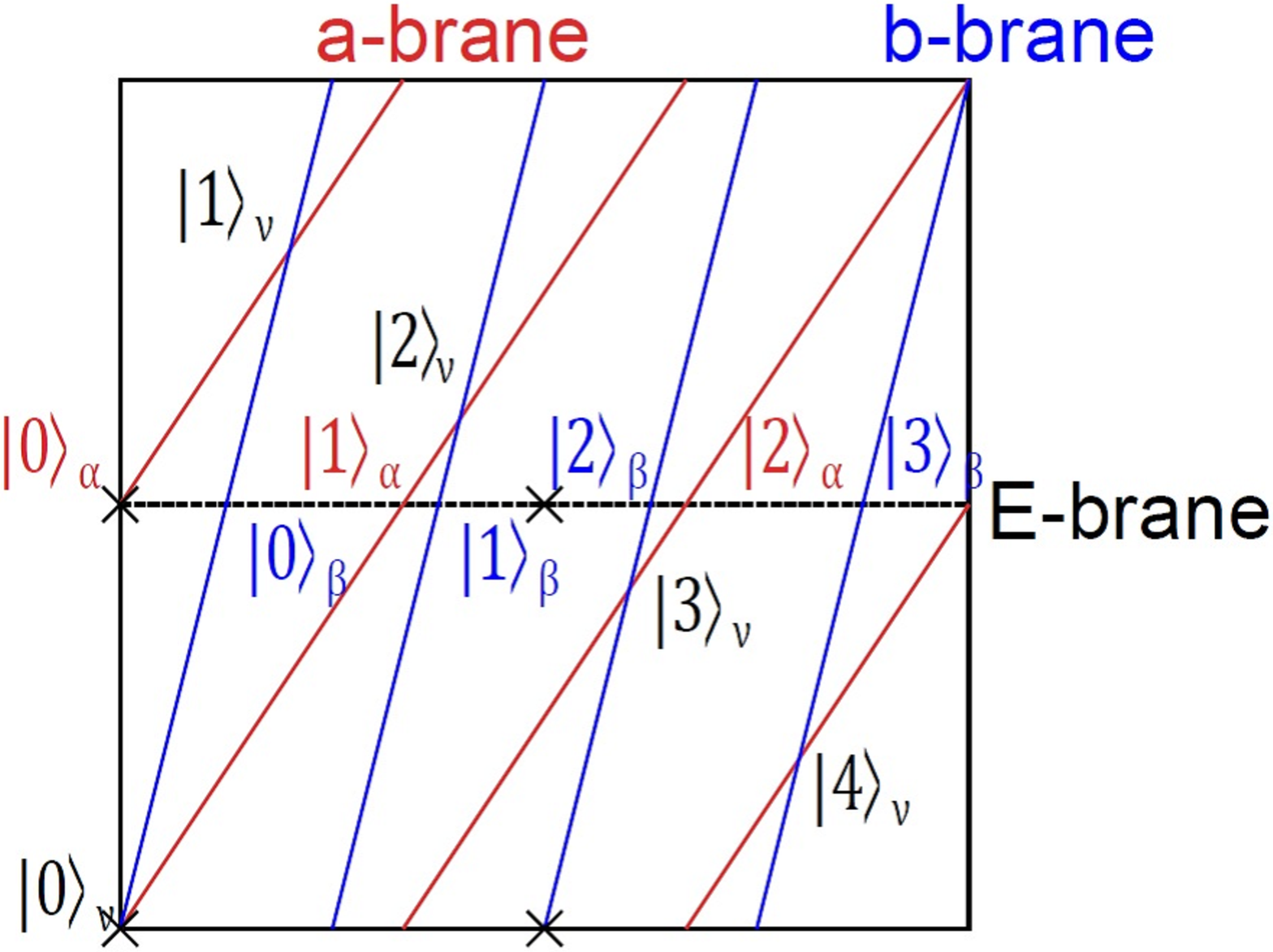,scale=0.25}
 \caption{The brane configuration on the first torus of Model 2.
We omit the index number of other tori, e.g. $\ket{0}_\alpha$ denotes $\ket{000}_\alpha$, the others are similar.}
\label{fig:ex2}
\end{minipage}
\end{tabular}
\end{figure}

There are two geometric symmetries in Models 1-(I) and 1-(II).
In Model 1-(I) , the flavor structure originates from the second torus.
On the second torus, there is $\mathbb{Z}_{2}$ symmetry acting branes as exchanging a-brane and b-brane and exchanging two fixed points simultaneously.
This symmetry exchanges $N_0, \alpha_i$ and $N_2, \beta_i$, but does not change the areas of worldsheet instantons. 
Then, the 3-point couplings are the same (see fig.\ref{fig:ex1}).
In Model 1-(II), the flavor structure  originates from the first torus.
However, there are no differences between $N_1$ and $N_2$  and 
the system is invariant under exchanging $N_1$ and $N_2$.
Thus, we realize the  $\mathbb{Z}_2$ permutation symmetry and bimaximal mixing mass matrix.

On the other hand, in Model 2, such $\mathbb{Z}_2$ symmetries no longer remain.
The flavor structure originates from the second torus.
We can not exchange $N_1$ and $N_2$ or fixed points (see fig.\ref{fig:ex2}).
The only symmetry is $\mathbb{Z}_2:z\rightarrow -z$, which is orbifold projection.
We can distinguish each state from others.
Thus, we get general symmetric mass matrix.


It is phenomenologically interesting  that the $\mathbb{Z}_2$ symmetry remains in  Majorana masses of 
right-handed neutrinos.
Such a symmetry would be favorable to derive large mixing angle in the lepton sector.
Indeed, many studies have been done for realization of observed lepton mixing angles by 
assuming non-Abelian discrete flavor symmetries.
(See for review \cite{Altarelli:2010gt,Ishimori:2010au,King:2013eh}.)\footnote{
Moreover, it is found that non-Abelian discrete flavor symmetries appear at perturbative level 
in heterotic orbifold models \cite{Kobayashi:2004ya,Kobayashi:2006wq} 
and intersecting/magnetized brane models \cite{Abe:2009vi,BerasaluceGonzalez:2012vb}.}

\subsection{Numerical analysis}

In this subsection, we analyze the lepton flavor structure numerically for illustration.
To compute the mixing angles of leptons and mass splittings, we have to determine the Dirac mass matrix of leptons, the number of Higgs fields and vacuum expectation values (VEVs) of Higgs fields.
In our analysis, left-handed leptons $L_i$s are localized at the intersection points of D6$_a$-brane and D6$_c$-brane.
The up-type Higgs fields $H_{u}$s are localized at the intersection points of D6$_b$-brane and D6$_c$-brane.
All of branes are wrapping rigid cycles.
Indeed, there are many possibilities for the charged lepton sector.\footnote{See for three-generation models 
on magnetized orbifold models, e.g. \cite{Abe:2008sx, Abe:2015yva}, and those would correspond to T-duals of intersecting D-brane models.}
Results depend on its choice.
For just illustration of numerical study, 
we assume the Dirac masses of the charged leptons are diagonal and they are equal to the observed charged leptons masses, 
and the mixing angles of leptons are determined by the Dirac and Majorana masses of neutrinos.
They are determined by the area of torus, and the VEVs of Higgs fields.
We do not compute Higgs potential in this analysis and we use their VEVs as free parameters.

In table \ref{tab:numerical1}, we show two examples of  mixing angles and neutrino mass splittings given by D-brane instanton effects.
In these examples, there are two Higgs doublets in the models.
Then, there are three parameters; the area of torus $A_1/\alpha'$, the ratio of Higgs VEVs $\langle H_{u1}\rangle/\langle H_{u0}\rangle$ and the scale of Majorana masses $M_s e^{-S_E}.$\footnote{More precisely, there is one more parameter, the overall scales of Higgs VEVs.
However, we can absorb it in the scale of Majorana masses and it does not affect our analysis.}

In the Example 1, the Majorana mass matrix is the same as Model 1-(I) in the previous subsection (\ref{eq:bimaximal2}).
We set the winding numbers as
\begin{equation}
\begin{split}
I_{ac}^1=1,~I_{ac}^2=5,~I_{ac}^3=1,\\
I_{bc}^1=1,~I_{bc}^2=2,~I_{bc}^3=1.
\end{split}
\end{equation}
The numbers of shared fixed points are $(1,1,1)_{ac}$ and $(1,2,1)_{bc}$.
In the example 1, we set the area of torus $A_2/\alpha'=0.6$ and the Higgs VEV ratio $\langle H_{u1}\rangle/\langle H_{u0}\rangle=0.2$.

In the Example 2, the Majorana mass matrix is the same as the explicit model 2 in the previous subsection.
We set the winding numbers as
\begin{equation}
\begin{split}
I_{ac}^1=4,~I_{ac}^2=1,~I_{ac}^3=1,\\
I_{bc}^1=2,~I_{bc}^2=1,~I_{bc}^3=1.
\end{split}
\end{equation}
The numbers of shared fixed points are $(2,1,1)_{ac}$ and $(1,2,1)_{bc}$.
In the example 2, we set the area of torus  $A_1/\alpha'=2.3$ and the Higgs VEV ratio $\langle H_{u1}\rangle/\langle H_{u0}\rangle=0.3$.

Table \ref{tab:numerical1} shows that we can realize the approximate values of the mixing angles and the mass splitting of neutrinos by E-branes, but there are the small deviations between theoretical and observed values.
These deviations may be caused by the assumptions of the diagonal charged leptons mass matrix.
Small deviations from the diagonal charged lepton mass matrix could explain the observed data.
At any rate, our purpose of this subsection is just an illustration of numerical study.

\begin{table}
\begin{center}
\begin{tabular}{c|cc|c} 
\hline
\hline
Observables & Example 1 & Example 2 & Observed values \\ \hline
$(m_{\nu_1}, m_{\nu_2}, m_{\nu_3})$[eV] & $(0.017,0.018,0.052)$ & $(0.000026,0.00011,0.051)$& $<2.0$\\
$|m_{\nu_2}^2 - m_{\nu_1}^2|$ [eV$^2$]& $7.8 \times 10^{-5}$ & $ 1.2 \times 10^{-8} $ & $7.62 \times 10^{-5}$ \\
$|m_{\nu_3}^2 - m_{\nu_2}^2|$ [eV$^2$]& $2.4 \times 10^{-3}$ & $ 2.6 \times 10^{-3} $ & $2.55 \times 10^{-3}$ \\
${\rm sin}^2\theta_{12}$ &  0.341 &  0.253 & 0.259 - 0.359 \\
${\rm sin}^2\theta_{23}$ &  0.758 &  0.827 & 0.380 - 0.628 \\
${\rm sin}^2\theta_{13}$ &  0.0212 &  0.0603 & 0.0176 - 0.0295 \\
\hline
\hline
\end{tabular}
\caption{Sample values of numerical analysis of the lepton flavor structure.
The observed values are quoted from \cite{Agashe:2014kda,Tortola:2012te}.
In our analysis, we assume the normal hierarchy.}
\label{tab:numerical1}
\end{center}
\end{table}

\section{Higgs $\mu$-term matrix}

\begin{figure}[thbp]
\centering
  \epsfig{file=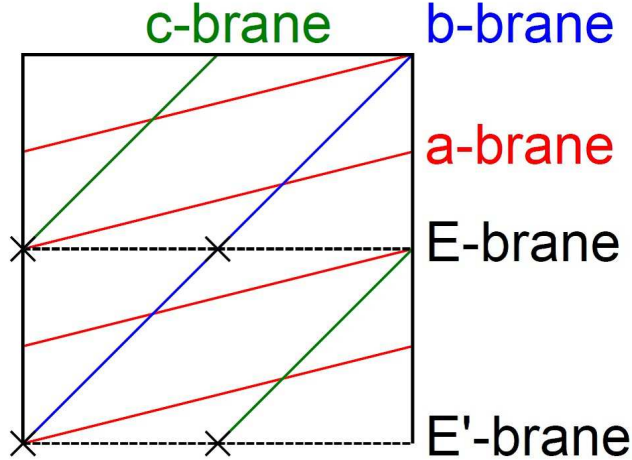,scale=0.25}
 \caption{The brane configuration on the first torus.
The intersection numbers of  E'-brane and other branes are the same as those of E-brane and other branes.}
\label{fig:mu}
\end{figure}

We can also obtain Higgs $\mu$-terms by E2-branes.
We consider $g$ pairs of Higgs fields $H_{u}$ and $H_{d}$.
We assume $H_u$s are localized at the intersection points of D6$_a$-brane and D6$_b$-brane and $H_d$s are localized at the intersection points of D6$_a$-brane and D6$_c$-brane.
The multiplicity of D6$_a$-brane is 2 and the others are 1.
To generate $\mu$-terms, the E-brane intersects other branes once, and there are three kinds of zero-modes $\alpha,\beta,\gamma$. 
$\alpha$ is localized at the E-a intersection point, $\beta$ is localized at the E-b  intersection point and $\gamma$ is at the E-c intersection point.
Then, we obtain the $\mu$-term,
\begin{equation}
\int \mathcal{D}^{2}\alpha \mathcal{D}\gamma \mathcal{D}\beta M_s e^{-S_{E}}e^{y_{i}^u  \alpha \cdot H_u^i \beta+y_{j}^d \alpha \cdot H_d^j \gamma}=M_s e^{-S_{E}} y_i^u y_j^d H_u^i \cdot H_d^j.
\end{equation}
This matrix shows that one instanton configuration can generate only rank-one $\mu$-term matrix.
Let us consider the model with two-pair Higgs fields for instance.
We study the model with $(1,1,1)_{ab,c},(0,0,0)_{bc},(1,1,1)_{Ea,b,c} $.
In this model, the intersection number can have the following solution,
\begin{equation}
\begin{split}
I_{ab}^{1}=I_{ac}^{1}=3,~~I_{ab}^{2}=I_{ac}^{2}=1,~~I_{ab}^{3}=I_{ac}^{3}=1,~I_{bc}^{i}=0~~&({\rm for}~\forall i \in \{ 1,2,3\}),\\
I_{Ex}^{1}=I_{Ex}^{2}=I_{Ex}^{3}=1~~&({\rm for}~\forall x\in\{a,b,c\}).
\end{split}
\end{equation}
Then, we obtain the  rank-one $\mu$-term matrix  similarly
\begin{eqnarray}
\mu_{ij} \propto \left(
\begin{array}{cc}
y_{0}^u y_{0}^d& y_{0}^u y_{1}^d\\
y_{1}^u y_{0}^d& y_{1}^u y_{1}^d\\
\end{array}
\right),
\end{eqnarray}
where $y^{u(d)}_i$ is the 3-point coupling of $i$-th Higgs field $H_{u(d)}^i$ and zero-modes on the first torus.
Because those on the other torus is common, we omit indices of torus. 
If there is nothing to obstruct, we also take into account the contribution of E-brane (E'-brane in fig.\ref{fig:mu}) wrapping the other fixed points on the first torus and having the same winding number.
Thus, we get rank-two symmetric $\mu$-term matrix,
\begin{eqnarray}
\mu_{ij} \propto \left(
\begin{array}{cc}
2y_{0}^u y_{0}^d& y_{0}^u y_{1}^d +y_{1}^u y_{0}^d\\
y_{0}^u y_{1}^d +y_{1}^u y_{0}^d& 2y_{1}^u y_{1}^d\\
\end{array}
\right).
\end{eqnarray}
This expression is inherited from the symmetry of the configurations of D-branes too.
The configuration is invariant under exchanging the fixed points and b,c-branes simultaneously.
Then, such an expression appears.
The rank of $\mu$-term matrix depends on the number of allowed configurations of E-branes.
In this example, there are two configurations and we get rank-two matrix.
If the  number of Higgs pairs is larger than two,  we can not realize the full rank $\mu$-term by one type of E-branes.
However, there are some possibilities that the configuration of rigid E-branes wrapping another bulk cycle can also induce some corrections for $\mu$-terms.
The suppression factor of E-brane is $e^{-S_{{\rm DBI}}}$.
Although the leading order E-brane induces high-scale $\mu$-terms, e.g. $\mathcal{O}(10^{10-15})$ GeV, 
the next to leading order term may be more suppressed.
In addition to that, the $\mu$-terms induced by E-branes is a product of the 3-point couplings.
The 3-point couplings are suppressed by worldsheet area.
If the compactification scale is large enough, Yukawa couplings are suppressed.
We may realize the low scale $\mu$-term naturally like this and solve the ``$\mu$-problem".

\section{Conclusion and discussion}

We have studied the effects of  E-branes on $\mathbb{Z}_2\times\mathbb{Z}_2'$ toroidal orbifold.
On this orbifold, there are rigid cycles, which can not move away from the fixed points and have no position moduli.
D-brane instanton wrapping such cycles can induce non-perturbative superpotential.

In this paper, we have concentrated on  Majorana masses of right-handed neutrinos as well as  Higgs $\mu$-terms.
We have computed the general form of Majorana neutrino masses and computed them explicitly in concrete models.
As a result, we have realized the bimaximal mixing Majorana mass matrix, 
although we can also obtain not so symmetric one.
This symmetry originates from geometric configurations of branes.
From phenomenological point of view, such Majorana mass matrix is interesting 
because one could derive the observed large  mixing angles by using such a symmetry.
In fact, we find some examples roughly fitting the mixing angles and neutrino mass splittings.

We also computed the Higgs $\mu$-term matrix in an explicit model and obtain rank-one and rank-two $\mu$-term matrices.
This is because one D-brane instanton can only make rank-one matrix and rigid E-brane wrapping the same bulk cycle has at most two configurations. 
However, we can consider another E-brane configuration and may derive the full rank $\mu$-term matrix.
Such configuration may induce suppressed term and explain the electroweak scale.
We would study it elsewhere.

It is important to study other compactifications.
If D-brane configurations including D-brane instantons have geometrical symmetries such as 
$\mathbb{Z}_2$ symmetry, Majorana neutrino masses would respect such symmetries and 
lead to phenomenologically interesting results even in more complicated compactifications.

\subsection*{Acknowledgement}

 T.K. and S.U. are supported in part by
the Grant-in-Aid for Scientific Research No.~25400252 and No.~15J02107 from the Ministry of Education,
Culture, Sports, Science and Technology  in Japan.

\appendix 

\section{Yukawa couplings of zero-modes and neutrinos}
\label{app:yukawa}

In this appendix, we compute Yukawa couplings of zero-modes and neutrinos explicitly in a model.
We study Model 1-(I) in Section \ref{subsec:ex}.
In this model, there are two branes (a-brane and b-brane) and one E-brane.
Their intersection numbers are
\begin{equation}
(I_{ab}^{1},I_{ab}^{2},I_{ab}^{3})=(1,6,2),~(I_{Ea}^{1},I_{Ea}^{2},I_{Ea}^{3})=(I_{Eb}^{1},I_{Eb}^{2},I_{Eb}^{3})=(1,3,1).
\end{equation}
The flavor structure of neutrinos and zero-modes originates from the second torus and we concentrate on it.
The brane configuration on the second torus is shown in fig.\ref{fig:ex1}.

Before computing Yukawa couplings, it is worth considering the geometrical symmetry of this configuration.
For example, since the area of the triangle whose vertices are $\ket{0}_\nu$, $\ket{0}_\alpha$ and $\ket{2}_\beta$ is the same as that of the triangle surrounded by $\ket{1}_\nu$, $\ket{1}_\alpha$ and $\ket{2}_\beta$, $y_{002}$ is equal to $y_{112}$.
Similarly, we obtain three equations;
\begin{equation}
\begin{split}
y_{001}=y_{112}=y_{210}=y_{320}=y_{421}=y_{501}=y_{0},\\
y_{011}=y_{100}=y_{222}=y_{311}=y_{400}=y_{522}=y_{1},\\
y_{020}=y_{121}=y_{201}=y_{302}=y_{412}=y_{510}=y_{2},
\end{split}
\end{equation}
and other couplings are forbidden.
Each coupling is written by a theta function.
\begin{equation}
y_{0}= C \vartheta \left[
\begin{array}{c}
\frac{1}{12} \\
0  \\
\end{array}
\right] \left( 0,\frac{i 6\pi A}{\alpha'} \right) ,
\end{equation}
\begin{equation}
y_{1}= C \vartheta \left[
\begin{array}{c}
\frac{3}{12} \\
0  \\
\end{array}
\right] \left( 0,\frac{i 6\pi A}{\alpha'} \right) ,
\end{equation}
\begin{equation}
y_{2}= C \vartheta \left[
\begin{array}{c}
\frac{5}{12} \\
0  \\
\end{array}
\right] \left( 0,\frac{i 6\pi A}{\alpha'} \right) ,
\end{equation}
where $C$ is the common factor of quantum corrections and $A$ is the area of the second torus.
The Yukawa couplings of the invariant states are the linear combinations of them.
We can express these couplings as
\begin{equation}
\begin{split}
d_0^{00}=0,~d_0^{01}=\sqrt{2} y_2,~d_0^{10}=\sqrt{2} y_0,~d_0^{11}=y_1,\\
d_1^{00}=2 y_1,~d_1^{01}=0,~d_1^{10}=0,~d_1^{11}=y_0+y_2,\\
d_2^{00}=0,~d_2^{01}=\sqrt{2} y_0,~d_2^{10}=\sqrt{2} y_2,~d_2^{11}=y_1,\\
\end{split}
\end{equation}
where $d_a^{ij}$ denote the Yukawa coupling of neutrino $N_a$ and zero-modes $\alpha_i$, $\beta_j$.

Finally, we obtain the interaction terms of eq.(\ref{eq:interact}) and the Majorana mass matrix of eq.(\ref{eq:bimaximal2}).

\end{document}